\documentclass[sigplan,twocolumn,accept]{acmart}
\renewcommand\footnotetextcopyrightpermission[1]{}
\AtBeginDocument{%
  }

\usepackage{listings}
\usepackage{array}
\usepackage{booktabs}
\usepackage{makecell}
\usepackage{subfig} 

\usepackage{algorithm}
\usepackage{algpseudocode}

\usepackage{tabularx}

\newcommand{\F}[1]{\mathsf{#1}}

\newcommand{\fstar}{F$^{\star}$\xspace}

\lstdefinestyle{leanstyle}{language=lean}

\usepackage{color}
\definecolor{keywordcolor}{rgb}{0.7, 0.1, 0.1}   
\definecolor{tacticcolor}{rgb}{0.0, 0.1, 0.6}    
\definecolor{commentcolor}{rgb}{0.4, 0.4, 0.4}   
\definecolor{symbolcolor}{rgb}{0.0, 0.1, 0.6}    
\definecolor{sortcolor}{rgb}{0.1, 0.5, 0.1}      
\definecolor{attributecolor}{rgb}{0.7, 0.1, 0.1} 

\usepackage{xcolor}
\usepackage{hyperref}
\usepackage{xspace}

\begin{document}

\title{\sal: Multi-modal Verification of Replicated Data Types}


\author{Pranav Ramesh}
\affiliation{%
  \institution{Indian Institute of Technology, Madras}
  \city{Chennai}
  \country{India}}
\email{cs22b015@smail.iitm.ac.in}

\author{Vimala Soundarapandian}
\affiliation{%
  \institution{Indian Institute of Technology, Madras}
  \city{Chennai}
  \country{India}}
\email{cs19d750@smail.iitm.ac.in}

\author{KC Sivaramakrishnan}
\affiliation{%
  \institution{Indian Institute of Technology, Madras}
  \city{Chennai}
  \country{India}}
\email{kcsrk@cse.iitm.ac.in}

\renewcommand{\shortauthors}{Ramesh et al.}
\newcommand{\sal}{\textsc{Sal}\xspace}

\begin{abstract}
Designing correct replicated data types (RDTs) is challenging because replicas evolve independently and must be merged while preserving application intent. A promising approach is correct-by-construction development in a proof-oriented programming language such as \fstar, Dafny and Lean, where desired correctness guarantees are specified and checked as the RDTs are implemented. Recent work Neem~\cite{Neem} proposes the use of replication-aware linearizability (RA linearizability)~\cite{WangRALin} as the correctness condition for state-based CRDTs and mergeable replicated data types (MRDTs), with automation in the SMT-aided, proof-oriented programming language \fstar. However, SMT-centric workflows can be opaque when automation fails to discharge a verification condition (VC), and they enlarge the trusted computing base (TCB).

We present \sal, a multi-modal workflow to design and verify state-based CRDTs and MRDTs in Lean. \sal combines (i) kernel-checkable automation with proof reconstruction, (ii) SMT-aided automation when needed, and (iii) AI-assisted interactive theorem proving for remaining proof obligations. When a verification condition is shown to be invalid, we leverage Lean's property-based testing to automatically generate and visualize counterexamples, helping developers debug incorrect specifications or implementations. We report on our experience verifying a suite of 13 CRDTs and MRDTs with \sal: 69\% of verification conditions are discharged by kernel-verified automation without SMT, and counterexamples automatically expose subtle bugs such as the well-known enable-wins flag anomaly. The codebase for \sal is open-sourced, and is available at \href{https://github.com/fplaunchpad/sal}{https://github.com/fplaunchpad/sal}.
\end{abstract}

\ccsdesc[500]{Software and its engineering~Software verification and validation}
\ccsdesc[500]{Computing methodologies~Distributed algorithms}
\keywords{CRDT, Verification, Lean, Multi-Modal Proofs, Counterexample Generation}


\maketitle

\section{Introduction}

Local-first collaboration tools allow users to continue working while offline, synchronizing in the background when connectivity returns~\citep{localfirst}. This programming model requires replicated data types whose states can evolve independently at each replica and later be merged without violating application intent. Conflict-Free Replicated Data Types (CRDTs)~\citep{CRDTs} and Mergeable Replicated Data Types (MRDTs)~\cite{MRDT} are widely used to implement such replicated state. However, designing correct RDTs is subtle. Even well-known designs such as Replicated Growable Arrays (RGA)~\cite{kleppmanRGA} have had serious anomalies discovered after publication\footnote{\url{https://martin.kleppmann.com/2019/03/25/papoc-interleaving-anomalies.html\#errata}}.

Replication-aware (RA) linearizability~\cite{WangRALin} provides a principled correctness condition for CRDTs, relating replica executions to a sequential explanation of updates. Prior work, Neem~\cite{Neem}, showed that RA linearizability can be extended to MRDTs and also be reduced to a set of 24 verification conditions (VCs) amenable to automation. Neem implements this approach in \fstar, an SMT-aided, proof-oriented programming language. Developers implement RDTs in \fstar and leverage SMT automation to discharge the RA-linearizability VCs automatically. In practice, however, SMT-centric verification workflows can be difficult to iterate on. When automation fails, developers often get little actionable information beyond an unproved VC, and debugging incorrect implementations remains manual and time-consuming.

This debugging experience contrasts sharply with everyday software development, where programmers iterate quickly by writing code, running tests, inspecting failing traces, and refining implementations. Understanding why a particular VC fails requires a laborious ``proof-debugging'' workflow: developers add intermediate assertions (\texttt{assert}) or assumptions (\texttt{admit}) and progressively push them deeper to localize the failing reasoning step. When the root cause is an unexpected interaction between definitions and the SMT encoding, additional tuning (e.g., controlling unfolding or solver limits) may be needed. Moreover, SMT-aided proofs can be brittle: small changes may cause previously discharged VCs to fail. Relying on an external SMT solver also increases the trusted computing base.

We present \sal, a verification workflow for RA-linearizability in Lean that addresses these issues by making failures actionable. At the core of \sal is a staged Lean tactic that prioritizes proof reconstruction (e.g., simplifying goals and invoking \texttt{grind}~\cite{lean_grind}) and falls back to SMT-aided automation (\texttt{lean-blaster}~\cite{lean_blaster}) only when necessary. When a VC is false, \sal uses property-based testing (Plausible~\cite{plausible}) to automatically synthesize small counterexamples and a ProofWidgets~\cite{ayers_et_al:LIPIcs.ITP.2021.4}-based visualizer to render the corresponding execution trace, enabling a test-like debugging loop for RDT verification. In our evaluation over a suite of CRDTs and MRDTs, \sal discharges most VCs without SMT and automatically rediscovers subtle bugs such as the well-known enable-wins flag anomaly (Table~\ref{tab:results_1}).

This paper makes the following contributions:
\begin{enumerate}
    \item A Lean formalization of RA-linearizability VCs for a suite of state-based CRDTs and MRDTs.
    \item A counterexample-generation and visualization workflow for failing VCs, based on property-based testing in Lean.
    \item \sal, a custom tactic that attempts proof reconstruction-based automation first, falls back to SMT-aided automation, and uses AI-assisted interactive theorem proving as a last resort.
    \item An evaluation across a suite of CRDTs and MRDTs, including cases where counterexamples expose subtle bugs (Table~\ref{tab:results_1}).
\end{enumerate}

\section{Background}
\label{sec:background}

We assume familiarity with replicated data types and focus on the specific
models and specifications used in this paper. State-based CRDTs reconcile
replicas via a deterministic two-way merge $\mu(v_1, v_2)$; a common sufficient
condition for convergence is that the merge function $\mu$ is the join of a
semilattice.

Mergeable Replicated Data Types (MRDTs)~\cite{MRDT} avoid embedding causal
metadata in the state of the RDT. Instead, they assume a versioned storage model
(e.g., Git-like histories) that can provide the lowest common ancestor (LCA) for
a three-way merge. This interface often yields compact implementations: for
example, a counter MRDT can be $O(1)$, whereas state-based counter CRDTs require
$\Omega(n)$ space in the number of replicas~\cite{Burckhardt}.

Formally, an MRDT implementation for a data type $\tau$~\citep{Neem} is a tuple
$\mathcal{D}_{\tau} = \langle \Sigma, \sigma_0, \F{do}, \F{merge}, \F{rc} \rangle$, where:
\begin{itemize}
	\item $\Sigma$ is the set of states, $\sigma_0 \in \Sigma$ is the initial state.
    \item $\F{do}$ : $\Sigma \times \mathcal{T} \times \mathcal{R} \times O_{\tau} \rightarrow \Sigma$ implements update operations parameterized by timestamp in $\mathcal{T}$, replica id in $\mathcal{R}$ and operations in $O_{\tau}$.
	\item $\F{merge}$ : $\Sigma \times \Sigma \times \Sigma \rightarrow \Sigma$ is a three-way merge function.
	\item $\F{rc} \subseteq O_{\tau} \times O_{\tau}$ is the conflict resolution policy to be followed for concurrent \emph{conflicting} update operations. The relation is interpreted as a partial order where $(o_1, o_2) \in \F{rc}$ means that $o_1$ is ordered before $o_2$.
\end{itemize}

For example, the increment-only counter MRDT is defined as follows:

\begin{itemize}
    \item $\Sigma = \mathbb{N}$ with $\sigma_0 = 0$
    \item $O = \{\F{inc}\}$
    \item $\F{do}(\sigma,\_,\_,\F{inc}) = \sigma+1$
    \item $\F{merge}(\sigma_{lca},\sigma_1,\sigma_2) = \sigma_{lca} + (\sigma_{1} - \sigma_{lca}) + (\sigma_2 - \sigma_{lca})$ -- the merged state is the sum of the states of the LCA and the difference between the LCA and the two versions.
    \item $\F{rc} = \emptyset$ -- no conflicting operations.
\end{itemize}
An observed-removed set (OR-set) MRDT~\cite{MRDT} is defined as follows:

\begin{itemize}
    \item $\Sigma = \mathcal{P}(\mathcal{T} \times E)$ where $E$ is the set of elements with $\sigma_0 = \emptyset$
    \item $O = \{\F{add}_e, \F{rem}_e | e \in E\}$
    \item $\F{do}(\sigma,t,\_,\F{add}_e) = \sigma \cup \{(e,t)\}$
    \item $\F{do}(\sigma,\_,\_,\F{rem}_e) = \sigma \setminus \{(e,i) | (e,i) \in \sigma\}$
    \item $\F{merge}(\sigma_{lca},\sigma_1,\sigma_2) = (\sigma_{lca} \cap \sigma_1 \cap \sigma_2) \cup (\sigma_1 \setminus \sigma_{lca}) \cup (\sigma_2 \setminus \sigma_{lca})$ -- the merged state contains elements common to all three versions (unchanged elements remain; deleted elements are not included), as well as elements added in either version since the LCA.
    \item $\F{rc} = \{(\F{rem}_e,\F{add}_e) | e \in E\}$ -- remove is ordered before add for the same element, and hence, adds win over concurrent removes.
\end{itemize}

We verify RDT correctness using replication-aware (RA) linearizability~\cite{WangRALin}, which makes merge semantics explicit while retaining a linearizability-style reading. Neem~\cite{Neem} reduces RA linearizability for MRDTs and state-based CRDTs to a finite set of verification conditions (VCs) over $\F{do}$, $\F{merge}$, and $\F{rc}$. \sal focuses on discharging these VCs effectively and on producing actionable feedback when automation fails.

Lean 4~\cite{lean_whitepaper} is a theorem prover and functional programming language with a small trusted kernel; proofs can be checked as proof terms by the kernel. Lean's metaprogramming support enables custom tactics and domain-specific automation, which is useful for VC-heavy developments like ours.

Lean offers several automation techniques with different tradeoffs. We rely on the following:
\begin{itemize}
    \item \texttt{dsimp} and \texttt{aesop}: simplification and proof search for routine goals.
    \item \texttt{grind}~\cite{lean_grind}: SMT-style automation with proof reconstruction, producing kernel-checkable proof terms.
    \item \texttt{lean-blaster}~\cite{lean_blaster}: an SMT backend (Z3) that is effective on many VCs (notably those with lambdas), but without proof reconstruction; it therefore enlarges the trusted computing base.
\end{itemize}

\sal stages automation accordingly: we first attempt reconstructible automation and only fall back to SMT when needed.

\section{\sal framework}

In this section, we describe the \sal multi-modal verification framework for RA-linearizability in Lean. We start with the challenge of designing data structures conducive to automated verification in Lean.

\subsection{Data structures for automated verification}
\label{sec:data_structures}

Many RDTs rely fundamentally on sets and maps. For example, the OR-set MRDT maintains a set of timestamped elements; map-based MRDTs and CRDTs require reasoning about extensional equality of mappings. However, Lean's standard library provides data structures optimized for different purposes: the mathematical $\F{Set}$ type uses propositions ($\alpha \to \F{Prop}$) suitable for manual proofs, while computational maps like $\F{RBMap}$ and $\F{HashMap}$ emphasize efficient iteration but complicate extensional reasoning.

For automated verification, we use decidable representations that tools like \texttt{grind} can reason about effectively. We therefore implement custom set and map interfaces inspired by \fstar's verification-oriented designs. Our sets use boolean-valued membership functions rather than propositions, since they typically contain datatypes where equality is decidable.

\begin{lstlisting}[language=lean]
abbrev set (a:Type) [DecidableEq a] := a → Bool
\end{lstlisting}

This representation is decidable by construction: since membership returns a boolean, automation can directly compute and compare set operations without requiring proof objects to reason about decidability. We require $\F{DecidableEq}$ for element types, ensuring all operations remain computable. This is a natural condition which is satisfied across the RDTs we verify with this framework.

Similarly, our map interface makes the domain explicit to enable extensional reasoning:

\begin{lstlisting}[language=lean]
structure map (key:Type) [DecidableEq key] (value:Type) where
  mappings: key → value
  domain: set key
\end{lstlisting}

Two maps are equal when their domains and mappings agree. Since the mapping is restricted to the domain $\F{key}$ which supports decidable equality, the map supports extensional equality.

Critically, we annotate all definitions and lemmas with $\F{@[simp, grind]}$ attributes and provide $\F{grind\_pattern}$ hints. This builds a domain-specific rewrite database: when \texttt{grind} encounters goals involving set membership, union, or map selection, it automatically applies the appropriate lemmas without manual guidance. These annotations trade the generality of Lean's standard library for automation-friendliness, enabling \texttt{grind} to discharge most VCs involving sets and maps without SMT assistance (Table~\ref{tab:results_1}). This is analogous to SMT patterns in \fstar.

\subsection{Counterexample generation in Lean}

When a VC fails during verification, understanding why it failed is critical for debugging. In SMT-centric workflows like \fstar, a failed VC provides little actionable feedback: developers must manually construct execution traces, add intermediate assertions, and progressively narrow down the source of the failure. This process is labor-intensive and requires significant expertise. In contrast, \sal leverages Lean's property-based testing framework, Plausible~\cite{plausible}, to automatically generate concrete counterexamples when VCs fail, transforming opaque proof failures into tangible test cases that developers can inspect and debug.

We demonstrate this approach using the enable-wins flag MRDT, a shared boolean flag that represents a disabled or enabled state. The desired specification is that in the case of concurrent enable and disable operations, the enable wins. This is expressed as $\F{disable} \xrightarrow{\F{rc}} \F{enable}$, meaning that when we read a replica's state, the flag should be true if there exists an enable operation that is not causally preceded by a disable.

\begin{figure}
\begin{lstlisting}[style=leanstyle]
abbrev concrete_st := Int × Bool
inductive app_op_t : Type where
| Enable
| Disable
abbrev op_t:= ℕ × ℕ × app_op_t
  /-timestamp, rid, operation-/
def do_ (s:concrete_st) (o: op_t) : concrete_st
:= match o with
| (_, (_, .Enable)) => (Prod.fst s + 1, true)
| (_, (_, .Disable)) => (Prod.fst s, false)
def merge_flag (l a b: concrete_st) :=
  if Prod.snd a && Prod.snd b then true
  else if not (Prod.snd a) && not (Prod.snd b) then false
  else if Prod.snd a then Prod.fst a > Prod.fst l
  else Prod.fst b > Prod.fst l
def merge (l a b: concrete_st) : concrete_st
:= (Prod.fst a + Prod.fst b - Prod.fst l , merge_flag l a b)
\end{lstlisting}
\caption{Buggy enable-wins flag MRDT implementation}
\label{fig:ewflag-impl}
\end{figure}

\begin{figure}
    \centering
    \includegraphics[width=0.7\linewidth]{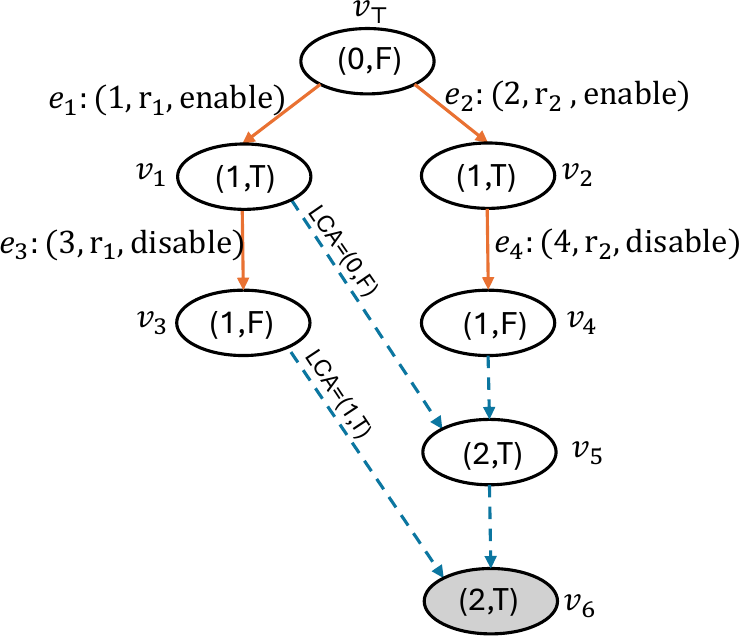}
    \caption{An enable-wins flag execution: both replicas see a disable at the end, yet merging produces $\F{(2,true)}$ at $v_6$, incorrectly reporting the flag as enabled.}
    \label{fig:ewflag-counter}
\end{figure}

Consider the enable-wins flag implementation in Figure~\ref{fig:ewflag-impl}, which tracks the number of concurrent enables using a counter and uses this counter to determine the flag's state after merging. This implementation contains a subtle bug. The bug manifests in executions where enable operations are followed by disables on each replica, yet the merged state incorrectly reports the flag as enabled. Figure~\ref{fig:ewflag-counter} shows such an execution: when merging $v_3$ and $v_5$ (with LCA $v_1$), the counter value of $v_5$ exceeds $v_1$, causing $\F{merge\_flag}$ to compute the new flag to be true in $v_6$. However, all enable events in this execution are subsequently disabled on their respective replicas, violating the enable-wins specification! When verifying the RA-linearizability VCs, one of the VCs fails. However, finding such counterexamples manually is challenging, particularly for more complex RDTs with larger state spaces.

To automate counterexample discovery, we leverage Plausible~\cite{plausible}, a property-based testing tool for Lean inspired by QuickCheck~\cite{quickcheck_paper}. Plausible requires that VCs be expressed as decidable propositions -- that is, properties that can be algorithmically evaluated to true or false. Once decidability is established, Plausible generates random test inputs and checks whether they satisfy the VC. When a violation is found, Plausible reports the failing input, such as:

\begin{lstlisting}[style=leanstyle]
postcondition violated for input
(disable, (enable, (enable, (disable, (0, false)))))
\end{lstlisting}

While this output identifies a counterexample, diagnosing the cause of the failure is difficult. To address this, we implement an execution trace visualizer using Lean’s ProofWidgets framework~\cite{ayers_et_al:LIPIcs.ITP.2021.4}. The visualizer instruments the $\F{do}$ and $\F{merge}$ operations to record intermediate states and operations, producing a step-by-step execution trace.

\sal, like Neem~\cite{Neem}, includes VCs that check whether concurrent executions can be RA-linearized to the same final state. This is achieved using \emph{bottom-up linearization}, where we peel off events in a bottom-up manner to construct a linearization order. In the failed VC, we attempt to prove:

\begin{align*}
\mu\big(\overbrace{e_1(0,false)}^{\textstyle v_1},\, \overbrace{e_3(e_1(0,false))}^{\textstyle v_3},\, \overbrace{e_1(e_4(e_2(0,false)))}^{\textstyle v_5}\big)
&=
\notag\\
e_3\big(\mu(\underbrace{e_1(0,false)}_{\textstyle v_1},\, \underbrace{e_1(0,false)}_{\textstyle v_1},\, \underbrace{e_1(e_4(e_2(0,false)))}_{\textstyle v_5})\big)
\end{align*}
where $\mu$ is the $\F{merge}$ operation, $e_1 \dots e_4$ are events, and each tuple $\F{(Int \times Bool)}$ constitutes a state. The state at $v_6$ in Figure~\ref{fig:ewflag-counter} is obtained by merging $v_3$ and $v_5$, with $v_1$ as LCA. This corresponds to the LHS of the VC. This state should match the state obtained by first merging $v_1$ and $v_5$, with $v_1$ as LCA, and then applying $e_3$, through bottom-up linearization.

\begin{figure}
    \centering
    \subfloat[]{
        \includegraphics[width=0.5\linewidth]{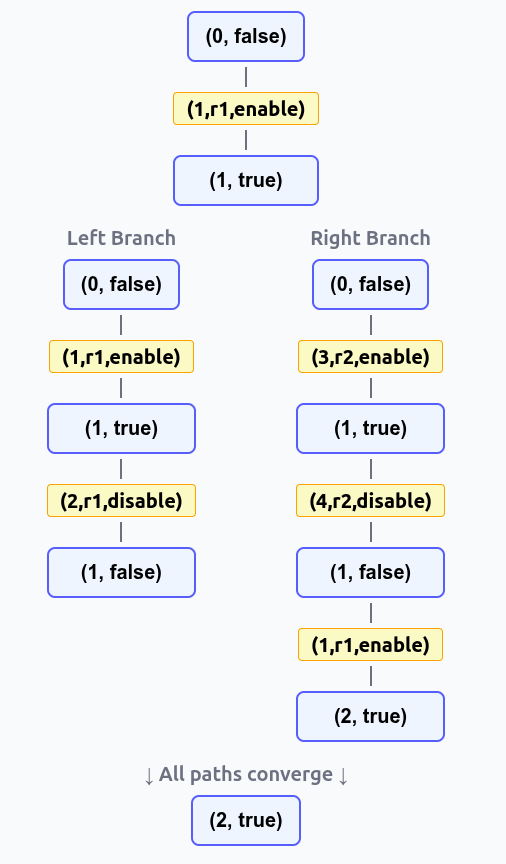}
        \label{fig:EN-wins-flag-LHS}
    }
    \subfloat[]{
        \includegraphics[width=0.5\linewidth]{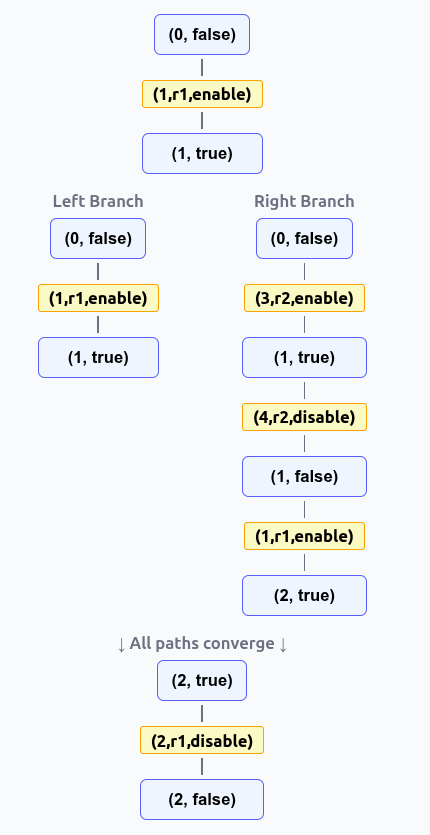}
        \label{fig:EN-wins-flag-RHS}
    }
    \caption{Visualization for the failed VC for the buggy enable-wins flag in Figure~\ref{fig:ewflag-impl}.}
    \label{fig:EN-wins-flag-combined}
\end{figure}

Figure~\ref{fig:EN-wins-flag-combined} illustrates the visualization for the
enable-wins flag that violates this VC. The states and operations are shown in
blue and yellow boxes, respectively, and correspond to $\F{concrete\_st}$ and $\F{op\_t}$, respectively, from Figure~\ref{fig:ewflag-impl}. The LCA trace and state is shown at the top. In both cases, the LCA state is $v_1$. The left
panel~(a) shows the execution trace for the LHS, which evaluates to
$\F{(2,true)}$, and the right panel~(b) shows the RHS, which evaluates to
$\F{(2,false)}$.  By comparing these traces, developers can quickly identify
that the bug arises from the counter-based merge logic failing to account for
subsequent disables. This visualization transforms the abstract VC failure into
a concrete debugging scenario, analogous to examining a failing unit test trace
in conventional software development.

The combination of automatic counterexample generation and interactive visualization significantly accelerates the debugging workflow. Instead of manually inspecting failed VCs and constructing hypothetical execution traces, developers receive concrete, visualizable counterexamples automatically. This approach also complements the multi-modal proof strategy: when \texttt{grind} or \texttt{lean-blaster} fail to discharge a VC, Plausible can quickly determine whether the failure stems from an incorrect implementation (producing a counterexample) or requires manual interactive proof (when no counterexample exists).

\subsection{Visualizing functional sets}

While the enable-wins flag uses simple concrete types (integers and booleans), many RDTs are set-based. Our custom sets (Section~\ref{sec:data_structures}), designed as functional predicates $\F{a} \to \F{Bool}$ for verification, pose a visualization challenge: they are abstract, infinite by nature, and do not support iteration or enumeration. Yet both counterexample debugging and general trace inspection require displaying actual set contents as concrete element lists.

\begin{figure}
    \centering
    \includegraphics[width=1\linewidth]{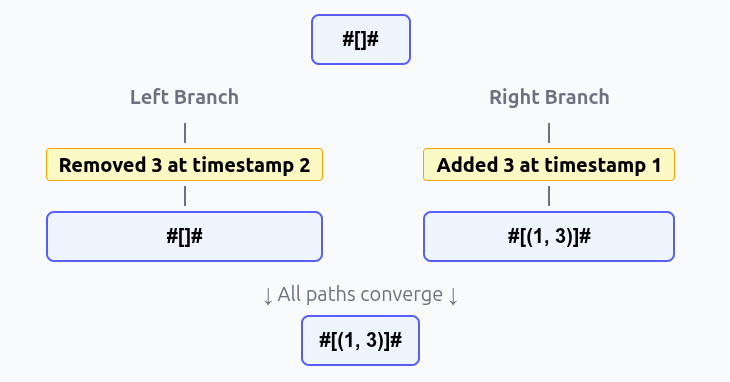}
    \caption{OR-set execution visualized using ProofWidgets}
    \label{fig:remove_wins_set_viz}
\end{figure}

To bridge this gap, we implement a universe tracking mechanism. During execution, we maintain a finite $\F{HashSet}$ of all elements that have been added or removed. When visualizing the set state, we only check membership for elements in this finite universe, which suffices to characterize the set's observable behavior. The implementation augments our abstract sets with a concrete universe:

\begin{lstlisting}[style=leanstyle]
structure set_with_universe (α: Type) [ToString α] [DecidableEq α] [Hashable α] where
  _set : set α
  _universe : HashSet α
\end{lstlisting}

Figure~\ref{fig:remove_wins_set_viz} demonstrates this approach on an OR-set execution with concurrent add and remove operations on element 3. The visualization shows concrete set states using the notation $\F{\#[(1, 3)]\#}$, where the tuple contains the timestamp and element. The left branch removes element 3 (resulting in the empty set $\F{\#[]\#}$), while the right branch adds element 3 (resulting in $\F{\#[(1, 3)]\#}$). Both branches converge to $\F{\#[(1, 3)]\#}$, confirming that adds win over concurrent removes as specified. Operations are displayed in yellow boxes, and users provide operation labels via format strings.

This mechanism works uniformly for both correct executions and counterexamples, enabling developers to inspect set-based RDT behavior regardless of whether they are debugging a failed VC or validating a correct implementation.

\subsection{Multi-modal proofs using the \sal tactic}

The \sal tactic orchestrates a staged proof strategy that adapts to VC complexity. As described in Section~\ref{sec:background}, Lean provides multiple automation techniques with different tradeoffs between power, trust, and performance. The \sal tactic attempts these approaches sequentially, prioritizing proof reconstruction before falling back to more powerful but less trustworthy methods.

The tactic proceeds in the following stages:
\begin{enumerate}
    \item \label{step1} \texttt{dsimp + grind (DG)} -- This combination applies simplification followed by \texttt{grind}'s SMT-style automation with proof reconstruction, producing kernel-checkable proof terms while maintaining the smallest TCB. We exclude \texttt{aesop}~\cite{aesop_paper} due to prohibitively high verification times on RDT VCs.

    \item \label{step2} \texttt{lean-blaster (LB)} -- When DG fails, \texttt{lean-blaster} encodes the goal to Z3. While more powerful for VCs with higher-order functions and lambdas, this sacrifices proof reconstruction and enlarges the TCB. We select it over Lean-SMT~\cite{lean_smt} and Lean-Auto~\cite{lean_auto} for its superior support for higher-order functions.

    \item \label{step3} \texttt{ITP} -- When both DG and LB fail, we discharge the remaining proof goals using Aristotle~\cite{achim2025aristotleimolevelautomatedtheorem}, an AI-assisted proof search tool. Aristotle generates kernel-checkable Lean proofs via extensive automated search. Unlike lean-blaster, the generated proofs are verified by Lean's kernel, preserving the TCB.
\end{enumerate}

In order to prevent runaway automation, the \sal tactic incorporates a heartbeat-based timeout mechanism that bounds the execution time of the first two stages. When a timeout expires, control returns to the user, allowing them to either increase the limit or proceed directly to interactive proving, either using Aristotle, or by hand using custom tactics.

\section{Evaluation}

\begin{table*}[htpb]
\centering
\caption{Comparing \fstar and Lean for verifying RA Linearizability}
\label{tab:fstar_lean_comp}

\begin{tabular}{p{0.14\textwidth} p{0.41\textwidth} p{0.41\textwidth}}
\toprule
\textbf{Criterion} & \textbf{\fstar} & \textbf{Lean} \\
\midrule
Automation &
Direct SMT solving via Z3. Automation often works out of the box at scale.
&
Multiple tactic-based tools (\texttt{aesop}, \texttt{grind}, \texttt{lean-blaster}), less effective on large proof goals.
\\
\midrule
Multi-Modal Proofs &
Limited tactic system; most proofs require SMT solver.
&
Rich tactic system allows combining automated and interactive proving.
\\
\midrule
Counterexamples &
No counterexample generation; reports only success or failure.
&
\texttt{Plausible} generates counterexamples for decidable properties.
\\
\midrule
Data Structures &
Rich set and map libraries designed for verification.
&
Mathematical data structures; custom sets and maps needed for automation (Section~\ref{sec:data_structures}).
\\
\midrule
Trustworthiness &
No proof reconstruction; relies on trusting Z3.
&
Proof reconstruction for most tools; kernel verifies proofs. Uncertified proofs marked \texttt{sorry}.
\\
\bottomrule
\end{tabular}
\end{table*}

\begin{table}[t]
\caption{Number of VCs discharged by the \sal tactic (total VCs = 24 per RDT). DG refers to $\F{dsimp + grind}$, and LB refers to $\F{lean-blaster}$. Time reports the average total wall-clock time for all stages including ITP via Aristotle. $^{\#}$Contains known bug; counterexample automatically generated (Section 3.2).}
\label{tab:results_1}
\centering
\footnotesize
\begin{tabular}{lcccc}
\toprule
\textbf{RDT} &
\textbf{DG} &
\textbf{LB} &
\textbf{ITP} &
\textbf{Time (s)} \\
\midrule
Increment-only counter MRDT    & 24 & 0  & 0 & 0.7  \\
PN-counter MRDT                & 24 & 0  & 0 & 2.7  \\
OR-set MRDT                    & 3  & 21 & 0 & 17.3 \\
Enable-wins flag MRDT $^{\#}$  & 9  & 14 & 0 & 34.2 \\
Efficient OR-set MRDT          & 2  & 22 & 0 & 14.4 \\
Grows-only set MRDT            & 24 & 0  & 0 & 0.3  \\
Grows-only map MRDT            & 22 & 0  & 2 & 25.8 \\
Replicated growable array MRDT & 15 & 9  & 0 & 3.9  \\
Multi-valued register MRDT     & 24 & 0  & 0 & 0.3  \\
Increment-only counter CRDT    & 24 & 0  & 0 & 1.5  \\
PN-counter CRDT                & 16 & 2  & 6 & 99.5 \\
Multi-valued register CRDT     & 24 & 0  & 0 & 0.7  \\
OR-set CRDT                    & 4  & 19 & 1 & 34.7 \\
\bottomrule
\end{tabular}
\end{table}

Table~\ref{tab:fstar_lean_comp} compares \fstar and Lean across several dimensions relevant to RDT verification. Lean's rich tactic system enables the multi-modal workflow central to \sal, while built-in counterexample generation (Plausible) and proof reconstruction support actionable debugging and reduced TCB. However, Lean's standard library data structures are designed for mathematical reasoning rather than automation, necessitating the custom sets and maps described in Section~\ref{sec:data_structures}.

Table~\ref{tab:results_1} shows the results of verifying 13 RDTs (311 VCs in total) using \sal. Across all benchmarks, \textbf{215 VCs (69.1\%) are discharged by dsimp+grind}, a lightweight tactic with proof reconstruction verified by Lean's kernel. An additional 87 VCs (28.0\%) require lean-blaster (SMT-based), and only 9 VCs (2.9\%) fall back to interactive proving. This demonstrates that the majority of RA-linearizability VCs can be verified without enlarging the TCB through SMT solvers.

We observe notable differences between CRDTs and MRDTs. MRDTs generally require less SMT than CRDTs due to their simpler three-way merge; for example, PN-counter MRDT needs no lean-blaster while PN-counter CRDT requires it for 2 VCs plus 6 interactive proofs. The 9 interactive proofs arise predominantly in CRDTs (7 of 9), where the merge function tends to involve more complex reasoning than three-way MRDT merges. Total verification times range from under a second for simple RDTs (e.g., grows-only set) to around 100 seconds for PN-counter CRDT, where the 6 ITP goals discharged by Aristotle dominate the total time.

The enable-wins flag MRDT demonstrates the value of counterexample generation. This implementation contains a known bug from prior work~\cite{Neem}; when the corresponding VC fails, Plausible automatically generates a concrete counterexample, which our ProofWidgets-based visualizer renders for inspection (Figure~\ref{fig:EN-wins-flag-combined}). This workflow transforms opaque VC failures into actionable debugging scenarios.

\section{Related work}

\textbf{Multi-modal verification.} Loom~\cite{loom_paper} is a recent work on multi-modal verification for data structures in Lean, using Dijkstra Monads to model effectful operations on mutable arrays and counters. \sal adapts this multi-modal approach to RDT verification in Lean, but leverages conventional tactics rather than monadic effects since RDTs are monotonic and non-mutable. Unlike Loom's focus on sequential data structures, \sal targets the unique challenges of replicated systems with merge operations and causal reasoning.

\textbf{CRDT verification.} Several approaches verify CRDT correctness. Isabelle/HOL has been used to verify strong eventual consistency~\cite{isabelle_verification}, but requires substantial manual proof effort. Neem~\cite{Neem} automates RA-linearizability verification in \fstar using SMT solvers, achieving high automation but at the cost of an enlarged TCB and opaque failures. Verifx~\cite{verifx_paper} provides a specialized language for RDTs with built-in verification, trading generality for domain-specific automation. It also supports automated counterexample generation in the context of RDT verification. \sal distinguishes itself through kernel-verified automation (69\% of VCs discharged by proof reconstruction), automated counterexample generation when verification fails, and interactive visualization of execution traces--capabilities not present together in prior CRDT verification tools.

\textbf{Counterexample generation.} Property-based testing tools like QuickCheck~\cite{quickcheck_paper}, QuickChick~\cite{quickchick} (Coq), and Plausible (Lean) generate test inputs for executable properties. \sal leverages Plausible for decidable VCs, but extends it with domain-specific visualization for RDT execution traces, including handling of functional set representations through universe tracking (Section 3.3). This bridges property testing and formal verification, transforming failed VCs into debuggable counterexamples.

\section{Conclusion}

We present \sal, a multi-modal verification framework for replicated data types in Lean that addresses key limitations of SMT-centric verification workflows. By staging automation--prioritizing kernel-verified proof reconstruction via \texttt{grind}, falling back to SMT-based \texttt{lean-blaster}, and using AI-assisted interactive theorem proving via Aristotle~\cite{achim2025aristotleimolevelautomatedtheorem} for remaining obligations--\sal achieves 69\% kernel-verified automation across 13 CRDTs and MRDTs while minimizing the trusted computing base. When automation fails, automated counterexample generation via Plausible and interactive visualization through ProofWidgets transform opaque proof failures into actionable debugging scenarios, as demonstrated by rediscovering the enable-wins flag bug.

Our experience reveals patterns in RDT verification: MRDTs generally require less SMT than CRDTs due to simpler three-way merge, and map-based reasoning remains more challenging than set-based reasoning for current Lean automation. Future work includes extending our custom data structure library to improve map automation, developing optimizations for the staged tactic based on VC patterns, and evaluating \sal on larger-scale RDT developments.

\bibliographystyle{ACM-Reference-Format}
\bibliography{sample-base}

\appendix

\end{document}